\def\gtrsim{\:\lower 0.4ex\hbox{$\stackrel{\scriptstyle >}
{\scriptstyle\sim}$}\:}
\def\sig{\:\lower 0.6ex\hbox{$\stackrel{\textstyle >}{\sim}$}\:}
\documentstyle[12pt,iopconf,epsf]{article}
\begin{document}
\title{Some nuclear physics aspects of core-collapse supernovae}

\author{Yong-Zhong Qian}

\affil{Physics Department, 161-33, California Institute of Technology,
Pasadena, CA 91125}

\beginabstract
Here I review some nuclear physics aspects of core-collapse supernovae
concerning neutrinos. Studies of neutrino emission and interactions
in supernovae are crucial to our understanding of the explosion mechanism, 
the heavy element nucleosynthesis, and pulsar proper motions. I discuss
the effects of reduced neutrino opacities in dense nuclear matter 
on supernova neutrino emission and their implications for
the delayed supernova explosion mechanism and the synthesis of neutron-rich
heavy elements.
I also discuss the effects of parity violation in weak interactions
on supernova neutrino emission and their implications for
pulsar proper motions.  
\endabstract

\section{Introduction}
In this paper I discuss some issues regarding core-collapse supernovae from
a nuclear physics viewpoint.

As we know, a supernova occurs when the core of a massive star collapses
into a compact neutron star with a mass of $M\sim 1\,M_\odot$ and a final
radius of $R\sim 10$ km. The gravitational binding energy of the final
neutron star is $E_B\sim GM^2/R\sim 3\times 10^{53}$ erg, with $G$ 
the gravitational constant. Due to the high temperatures
and densities encountered during the collapse, the most efficient way to
release this energy is to emit all six species of neutrinos and antineutrinos
mainly through electron-positron pair annihilation. By comparison, the total
kinetic energy of the debris ejected in the supernova explosion,
$\sim 10^{51}$ erg, is only
$\sim 1$\% of the total energy emitted in $\nu_e$ and $\bar\nu_e$. Exactly how
this $\sim1$\% of the neutrino energy is deposited in the matter above the
neutron star lies at the heart of the supernova explosion mechanism.

Furthermore, rotating magnetized neutron stars, commonly known as pulsars,
are observed to have large spatial velocities with an average value of
$v_{\rm kick}\approx 450$ km/s \cite{lyne}. 
If this velocity is the result of a kick imparted
to the neutron star at its birth, the momentum of this kick is
$P_{\rm kick}\sim 10^{41}$ g~cm/s. 
By comparison, the sum of the momentum magnitude
for all the neutrinos emitted in a supernova is $E_B/c\sim 10^{43}$
g~cm/s. Of course, if these neutrinos are emitted isotropically, the net
neutrino momentum is zero and no kick will be imparted to the neutron star.
However, if $\sim 1$\% of these neutrinos are emitted in a preferential
direction, then the resulting asymmetry in the neutrino momentum distribution
is sufficient to explain the observed average pulsar kick velocity.

Finally, supernovae play an essential role in the chemical evolution
of our universe. Many heavy elements, especially the very neutron-rich ones,
are believed to come from supernovae. A famous or infamous example is uranium,
which was used to create mankind's own version of a big explosion.
The synthesis of these heavy elements require certain physical conditions.
As we shall see, these conditions are determined by the neutrinos in the
supernova environment. 

Because neutrinos are of such great importance to the supernova process,
in the following discussions,
I will focus on the studies of supernova neutrino emission and on
their implications for the explosion mechanism, the synthesis of neutron-rich
heavy elements, and pulsar kick velocities.

\section{Neutrino opacities in dense nuclear matter and the supernova explosion
mechanism}
A relatively recent development in supernova studies is the discovery of
the delayed supernova explosion mechanism by Jim Wilson \cite{bethe}. 
This mechanism
can be briefly described as follows. A few tenth of a second after the
core collapse, the shock wave generated by the core bounce following the
collapse is stalled at $\sim 300$ km away from the neutron star. The material
between the neutron star and the stalled shock is mainly dissociated into
neutrons and protons due to the high temperatures (a few MeV) in this region.
As the neutrinos coming from the neutron star free-stream through this
material, a fraction of the $\nu_e$ and $\bar\nu_e$ are captured by the
nucleons, and their energy is deposited in the material through these 
capture reactions:
\begin{eqnarray}
\nu_e+n&\rightarrow& p+e^-,\\
\bar\nu_e+p&\rightarrow& n+e^+.
\end{eqnarray}
In other words, the material behind the shock is heated by the neutrinos.
If this neutrino heating is efficient enough, the stalled shock can be
revived to make a successful supernova explosion. Since this mechanism
operates after the shock fails to deliver a prompt explosion, it is called
the delayed supernova explosion mechanism.

The neutrino heating rate is approximately given by
\begin{equation}
\dot q\sim\phi_\nu\langle E_\nu\sigma_{\nu N}\rangle\propto
{L_\nu\over\langle E_\nu\rangle}\langle E_\nu^3\rangle,
\end{equation}
where $\phi_\nu$ is the neutrino flux, $E_\nu$ is the neutrino energy,
$\sigma_{\nu N}\propto E_\nu^2$ 
is the cross section for neutrino capture on nucleons,
$L_\nu$ is the neutrino luminosity, and where $\langle~\rangle$ denotes
the average over the neutrino energy spectrum. Therefore, the efficiency
of neutrino heating and the success of the delayed supernova explosion
mechanism depend crucially on the luminosities and energy spectra for
the $\nu_e$ and $\bar\nu_e$. In turn, these characteristics of supernova
neutrino emission are determined by the neutrino transport processes inside 
the neutron star, and thus depend on the neutrino opacities in the dense
nuclear medium.

The main opacity sources for $\nu_e$ and $\bar\nu_e$ are the same capture 
reactions on nucleons in Eqs. (1) and (2),
which are responsible for the neutrino heating above
the neutron star. However, in contrast to the situation above the neutron
star, the nucleons inside the compact neutron star are subject to frequent
interactions with other nucleons. As shown recently by a number of studies
\cite{sawyer,keil,reddy},
frequent interactions between nucleons tend to reduce the neutrino opacities
in the nuclear medium. With smaller neutrino opacities, the neutrinos can
get out of the neutron star faster. In addition, they can also decouple
from the neutron star matter at higher densities and temperatures. 
Consequently, the neutrino luminosities and average energies will increase,
whereas the duration of the neutrino emission will decrease. Therefore, one
expects that implementing these new neutrino opacities in the supernova
code may increase the efficiency of neutrino heating and make the delayed
explosion occur more easily. 

I should stress here that changing neutrino opacities is not as trivial as
tuning some free parameters. It also has important consequences for future
supernova neutrino detection. As we recall, the Kamiokande and IMB detectors
observed a total of 19 neutrino events from SN 1987A in the Large Magellanic 
Cloud \cite{kam,imb}. 
If we are lucky enough, the much larger super Kamiokande detector
can see several thousand neutrino events from a supernova in our Galaxy.
With such a supernova neutrino detection, we can certainly determine, for
example, the average $\bar\nu_e$ energy to better than 10\%, and therefore
can test the change of neutrino opacities due to the nuclear medium effects.
In addition, since these several thousand neutrino events will spread over
a period of about 10 s as in the case of SN 1987A, we can also learn a lot 
about the time evolution of supernova neutrino emission from such a detection.

\section{Time evolution of neutrino energy spectra and heavy element 
nucleosynthesis in supernovae}
The time evolution of the energy spectra for $\nu_e$ and $\bar\nu_e$ are
of particular interest to us. The solid curve in Fig. 1 
shows the evolutionary track of the
mean energies for $\nu_e$ and $\bar\nu_e$ calculated in Jim Wilson's
supernova model. The mean neutrino energy is defined as
\begin{equation}
\epsilon_\nu\equiv{\langle E_\nu^2\rangle\over\langle E_\nu\rangle}.
\end{equation}
The reason for this definition will become clear shortly. 
The open circles on the track in Fig. 1 indicate the time after the core
collapse from $t\approx 0$ s at the lower end to $t\approx 18$ s at the
upper end. Time increases along the track in intervals of approximately
1/3 s for $t\approx 0$--4 s and approximately 1 s for $t\approx 4$--18 s.
As we can see,
at $t< 1$ s after the core collapse, $\epsilon_{\nu_e}$ and 
$\epsilon_{\bar\nu_e}$ are comparable. However, at $t>2$ s, $\epsilon_{\nu_e}$
keeps decreasing, whereas $\epsilon_{\bar\nu_e}$ keeps increasing
with time. This time evolution can be understood as follows. First of all,
as mentioned previously, the main opacity sources for $\nu_e$ and $\bar\nu_e$ 
are the capture reactions on neutrons and protons in Eqs. (1) and (2), 
respectively. At $t<1$ s, 
although the neutron star has more neutrons than protons, the neutron and 
proton concentrations are still comparable. This means that the opacities
for $\nu_e$ and $\bar\nu_e$ are also comparable. Therefore, although the
$\bar\nu_e$ have a somewhat smaller opacity and correspondingly have a
somewhat larger mean energy, this mean energy $\epsilon_{\bar\nu_e}$
is not too different from $\epsilon_{\nu_e}$. However, as time goes on,
the net neutronization process of electron capture on protons inside the
neutron star increases the neutron concentration and decreases the proton
concentration. This results in increasing $\nu_e$ opacities and decreasing
$\bar\nu_e$ opacities. Consequently, $\epsilon_{\nu_e}$ decreases
whereas $\epsilon_{\bar\nu_e}$ increases with time.

\begin{figure}
\vskip 1cm
\centerline{\epsfxsize=6in\epsfbox{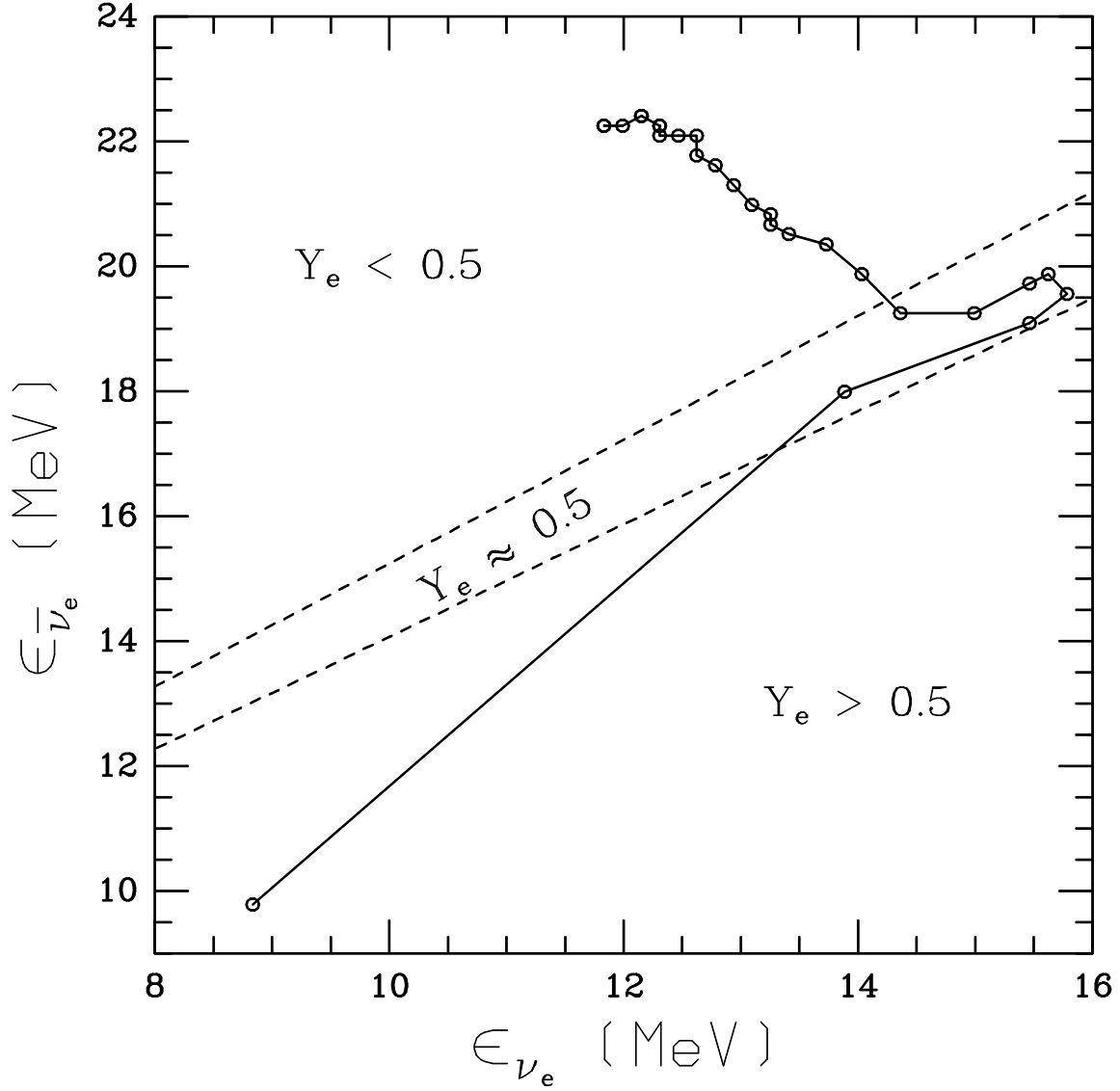}}
\caption{Time evolution of neutrino energy spectra and heavy element
nucleosynthesis in supernovae. The solid curve shows the evolutionary track 
of the mean neutrino energies
$\epsilon_{\bar\nu_e}$ and $\epsilon_{\nu_e}$ in Jim Wilson's supernova
model. The open circles on the track indicate the time after the core
collapse from $t\approx 0$ s at the lower end to $t\approx 18$ s at the
upper end. Time increases along the track in intervals of approximately
1/3 s for $t\approx 0$--4 s and approximately 1 s for $t\approx 4$--18 s.
Three regions, separated by the two dashed lines, correspond to values
of $\epsilon_{\bar\nu_e}$ and $\epsilon_{\nu_e}$ that would give $Y_e>0.5$,
$Y_e\approx 0.5$, and $Y_e<0.5$, respectively, in the supernova ejecta.
See text for detailed explanation.}
\end{figure}

The difference between $\epsilon_{\nu_e}$ and $\epsilon_{\bar\nu_e}$ is
crucial to the nucleosynthesis of neutron-rich heavy elements in supernovae.
This is because the neutrino capture reactions on nucleons above the neutron
star not only provide the heating for the revival of the stalled shock, they
also interconvert neutrons and protons, and therefore determine the 
neutron-to-proton ratio, or equivalently the electron fraction $Y_e$,
in the supernova ejecta. Clearly, in order to make neutron-rich heavy
elements, the ejecta have to be neutron-rich to start with. In other words,
$Y_e$ in the ejecta has to be less than 0.5. The neutron-to-proton ratio
in the ejecta is approximately given by the ratio of the neutron production
rate to the proton production rate \cite{qian1}. These rates are given by 
\begin{eqnarray}
\lambda_{\nu_en}&=&\phi_{\nu_e} \langle\sigma_{\nu_en}\rangle\propto
{L_{\nu_e}\over\langle E_{\nu_e}\rangle}\langle(E_{\nu_e}+\Delta)^2\rangle
\approx L_{\nu_e}(\epsilon_{\nu_e}+2\Delta),\\
\lambda_{\bar\nu_ep}&=&\phi_{\bar\nu_e} \langle\sigma_{\bar\nu_ep}
\rangle\propto{L_{\bar\nu_e}\over\langle E_{\bar\nu_e}\rangle}\langle
(E_{\bar\nu_e}-\Delta)^2\rangle\approx L_{\bar\nu_e}
(\epsilon_{\bar\nu_e}-2\Delta),
\end{eqnarray}
where we have taken into account the effects of the
neutron-proton mass difference $\Delta$ on the electron and positron
energies associated with the reactions in Eqs. (1) and (2).
With the mean neutrino energy defined in Eq. (4), the rates 
in Eqs. (5) and (6) are 
approximately proportional to the product of the neutrino luminosity and
the mean neutrino energy. However, because the neutron is heavier than the
proton by $\Delta \approx 1.3$ MeV, the conversion of a proton into a
neutron by the $\bar\nu_e$ is actually more difficult than the conversion
of a neutron into a proton by the $\nu_e$. As a result, for the same neutrino
luminosity, $\epsilon_{\bar\nu_e}$ has to exceed $\epsilon_{\nu_e}$ by 
about $4\Delta\approx 5.2$ MeV in order to drive the ejecta neutron-rich.

From the above discussion, we can identify three regions in Fig. 1 
according to the electron fraction $Y_e$ of the supernova ejecta \cite{qian2}.
If the ratio $L_{\bar\nu_e}/L_{\nu_e}$ is fixed, the mean neutrino energies
corresponding to $Y_e=0.5$, or equivalently
$\lambda_{\bar\nu_ep}=\lambda_{\nu_en}$, 
will lie on a single straight line in Fig. 1.
However, numerical supernova models show that 
$L_{\bar\nu_e}/L_{\nu_e}\approx1$--1.2. Consequently, the mean neutrino
energies corresponding to $Y_e\approx 0.5$ lie in a narrow region between
the two dashed lines in Fig. 1. The mean neutrino energies in the region
above both dashed lines 
correspond to $Y_e<0.5$, whereas those in the region below both dashed lines
correspond
to $Y_e>0.5$. From the evolutionary track of the mean neutrino energies 
in Jim Wilson's supernova model, we find that neutron-rich ejecta can be
obtained only at $t\gtrsim 3$ s. Consequently, the synthesis of neutron-rich
heavy elements can occur only at these late times. Of course, the onset
of this nucleosynthesis and the exact values of $Y_e$ in the ejecta
are again sensitive to the neutrino opacities inside the neutron star.
 
\section{Asymmetric supernova neutrino emission and pulsar kick velocities}
Finally, I discuss the connection between supernova neutrino emission
and pulsar kick velocities. As mentioned in the introduction, an asymmetry
of a few percent in the radiated neutrino energy can result in a kick 
momentum sufficient to explain the observed average pulsar velocity. One way
to induce such an asymmetry in neutrino emission is to take advantage of
parity violation in weak interactions. 

All neutrino species have intense
neutral-current scatterings on nucleons inside the neutron star:
\begin{equation}
\nu({\mbox{\bf k}})+N\rightarrow\nu({\mbox{\bf k}'})+N,
\end{equation}
where {\bf k} and $\mbox{\bf k}'$ 
are the incoming and outgoing neutrino momenta,
respectively. For 
illustration, we consider scatterings on neutrons only. If there is a
uniform magnetic field in the neutron star interior, the neutron spins will
be polarized in the direction opposite to that of the magnetic field.
Due to parity violation in weak interactions, the neutrino scattering cross 
section will then depend on the directions of both the incoming and outgoing
neutrinos with respect to the magnetic field direction $\hat{\bf B}$:
\begin{equation}
{\d\,\sigma_{\rm sc}\over \d\,\Omega_{\mbox{\bf k}'}}\propto 1+\beta_{\rm in}
\hat{\mbox{\bf k}}\cdot
\hat{\mbox{\bf B}}+\beta_{\rm out}\hat{\mbox{\bf k}}'\cdot\hat{\mbox{\bf B}}+
{\rm const.}\times
\hat{\mbox{\bf k}}\cdot\hat{\mbox{\bf k}}',
\end{equation}
where $\hat{\mbox{\bf k}}$ and $\hat{\mbox{\bf k}}'$ 
are the unit vectors along
{\bf k} and $\mbox{\bf k}'$, respectively, 
and $d\Omega_{\mbox{\bf k}'}$ is the differential
solid angle centered around $\mbox{\bf k}'$.
For scatterings on
neutrons, the cross section is more sensitive to the direction of the
outgoing neutrino. The preference for being scattered into a direction 
along the magnetic field is characterized by the neutron spin polarization
$P_n$:
\begin{equation}
\beta_{\rm out}\sim 
P_n\sim {\mu_nB\over T}\approx 6\times10^{-6}\left({B\over 10^{13}\ {\rm G}}
\right)\left({10\ {\rm MeV}\over T}\right),
\end{equation}
where $\mu_n$ is the magnitude of the neutron magnetic moment.
For typical conditions inside the neutron star, the directional asymmetry in 
a single neutrino scattering is only $\sim 10^{-5}$ for a magnetic field
of strength $B\sim 10^{13}$ G. 
However, because the neutrino mean free path $l$ for scattering
is $\sim 10^4$ times smaller than the neutron star radius $R$, these neutrinos
will be scattered many times before they are emitted from the neutron star.
As pointed out recently by Horowitz and Li \cite{horo}, 
the directional asymmetry in 
individual neutrino scatterings can accumulate through multiple scatterings
to give a much larger asymmetry in neutrino emission. These authors estimated
that the asymmetry $A$ in neutrino emission is
\begin{equation}
A\sim\alpha\beta_{\rm out}\left({R\over l}\right)\sim 0.006
\left({\alpha\over 0.1}\right)\left({B\over
10^{13}\ {\rm G}}\right)\left({10\ {\rm MeV}\over T}\right),
\end{equation}
where $\alpha\sim 0.1$ is a geometric factor. Consequently, they expected
that the observed average pulsar kick velocity could be produced by the
parity violation effects with a uniform neutron star magnetic field of 
strength $B\sim 10^{13}$ G.

However, while the cumulative effect of multiple scatterings on 
the neutrino emission asymmetry
given in Eq. (10) is qualitatively correct, the actual asymmetry in the
radiated neutrino energy can be obtained only from  
a systematic study of the parity
violation effects on neutrino transport in the context of neutron star
cooling. For the detailed results of such a study, see Ref. \cite{lai}.

\section{Conclusion}
To summarize, neutrinos play important roles in the supernova process,
and studies of their emission in supernovae are crucial to our understanding
of the explosion mechanism, the heavy element nucleosynthesis, 
and pulsar
kick velocities.

\section*{Acknowledgments}
Y.-Z. Qian was supported by the David W. Morrisroe Fellowship at Caltech.

\end{document}